# The effect of grain size on electrical transport and magnetic properties of $La_{0.9}Te_{0.1}MnO_3$

J. Yang[a], B. C. Zhao[a], R. L. Zhang[a], Y. Q. Ma[a], Z.G. Sheng[a], W. H. Song[a], Y. P. Sun[a,b],*

[a]Key Laboratory of Materials Physics, Institute of Solid State Physics, Chinese Academy of Sciences, Hefei, 230031, P. R. China

[b]National Laboratory of Solid State Microstructures, Nanjing University, Nanjing 210008, P. R. China

**Abstract**

The effect of grain size on structural, magnetic and transport properties in electron-doped manganites $La_{0.9}Te_{0.1}MnO_3$ has been investigated. All samples show a rhombohedral structure with the space group $R\bar{3}C$ at room temperature. It shows that the Mn-O-Mn bond angle decreases and the Mn-O bond length increases with the increase of grain size. All samples undergo paramagnetic (PM)-ferromagnetic (FM) phase transition and an interesting phenomenon that both magnetization and the Curie temperature $T_C$ decrease with increasing grain size is observed, which is suggested to mainly originate from the increase of the Mn-O bond length $d_{Mn-O}$. Additionally, $\rho$ obviously increases with decreasing grain size due to the increase of both the height and width of tunneling barriers with decreasing the grain size. The results indicate that both the intrinsic colossal magnetoresistance (CMR) and the extrinsic the extrinsic interfacial magnetoresistance (IMR) can be effectively tuned in $La_{0.9}Te_{0.1}MnO_3$ by changing grain size.

* Corresponding author. E-mail address: *ypsun@issp.ac.cn* (Y. P. Sun).

Mixed-valent manganites perovskites have attracted considerable attention in recent years because of the observation of CMR and more generally due to the unusually strong coupling between their lattice, spin, and charge degrees of freedom. Although the focus of interest has primarily rested with the hole-doped manganites $Ln_{1-x}A_xMnO_3$ (Ln = La-Tb, and A = Ca, Sr, Ba, Pb, etc.) due to their potential applications such as magnetic reading heads, field sensors and memories [1], naturally many researches have placed emphasis on electron-doped compounds such as $La_{1-x}Ce_xMnO_3$ [2-5], $La_{1-x}Zr_xMnO_3$ [6], $La_{2.3-x}Y_xCa_{0.7}Mn_2O_7$ [7], and $La_{1-x}Te_xMnO_3$ [8-10] due to the potential applications in spintronics. These investigations also suggest that the CMR behavior probably occur in the mixed-valent state of $Mn^{2+}/Mn^{3+}$. The basic physics in terms of Hund's rule coupling between $e_g$ electrons and $t_{2g}$ core electrons and Jahn-Teller (JT) effect due to $Mn^{3+}$ JT ions can operate in the electron-doped manganites as well.

From the viewpoint of future applications, the manganites with smaller grain size will be required. For this class of materials, lower sintered temperatures ($T_s$) will be necessary as the grain size increases with $T_s$. The effect of grain size on structural, magnetic and transport properties in hole-doped manganites has been extensively investigated [11-17]. All these researches indeed suggest that the magnetic properties are markedly affected by grain size. So it is of vital interest to investigate the effect of the grain size on the magnetoresistance and related transport properties because of the electrical transport properties of the manganites are closely linked to their magnetic properties. However, many controversial results have been reported concerning the influence of grain size on the magnetic properties of polycrystalline manganites. Sánchez et al. studied the effect of grain size in $La_{0.67}Ca_{0.33}MnO_3$ and found that both magnetization and the Curie temperature ($T_C$) decreased with decreasing grain size [11]. Hueso et al. also observed the similar rule and they proposed that it was mainly attributed to the presence of a nonmagnetic surface layer created by noncrystalline material that was important as the particle size decreases [12]. Zhang et al. investigated a structure-dependent change of magnetization in the manganite oxide $La_{0.85}Sr_{0.15}MnO_z$ and found that the contrary statements, i.e., both magnetization and $T_C$ decreases with increasing grain size [13]. They supposed that the contradiction could originate from the different doping level because the structure or magnetism was sensitive to the doping level as presented in [18]. To understand the essence of this issue, in this article, we will report our investigation of the effect of grain size on structural, magnetic and electrical transport property in the electron-doped manganites $La_{0.9}Te_{0.1}MnO_3$. We claim that our material is an electron-doped manganite based on simple valence arguments. In fact, X-ray photoemission spectroscopy (XPS) measurement revealed that the Te ions were in the tetravalent state and the manganese ions could be considered as in a mixture state of $Mn^{2+}$ and $Mn^{3+}$ for the compound $La_{1-x}Te_xMnO_3$ [8-10].

Nanometer samples of $La_{0.9}Te_{0.1}MnO_3$ were prepared by a citrate gel technique [19]. Stoichiometric amounts of high-purity $La_2O_3$, $TeO_2$, and Mn metal powders were dissolved in diluted nitric acid in which an excess of citric acid and ethylene glycol was added to make a metal complex. After all the reactants had completely dissolved, the solution was mixed and heated on a hot plate resulting in the formation of a gel. The gel was dried at 250 °C, and then preheated to 600 °C to remove the remaining organic and to decompose the nitrates of the gel. The resulting powder was pressed into pellets, each of which was sintered at a different temperature ($T_s$ = 800 °C, 900 °C and 1000 °C) in air. In order to obtain a polycrystalline reference pattern, ceramic samples of $La_{0.9}Te_{0.1}MnO_3$ were prepared by the conventional solid-state reaction method in air. Appropriate proportions of high-purity $La_2O_3$, $TeO_2$, and $MnO_2$ powders were thoroughly mixed according to the desired stoichiometry, and then calcined at 700 °C for 24 h. The powders obtained were ground, pelletized, and sintered at 1050 °C for 96 h with three intermediate grindings, and finally, the furnace was cooled down to room temperature. For the sake of description, the samples are labeled by sample A, B, C and D corresponding to their sintering temperature, i.e., 800 °C, 900 °C, 1000 °C and 1050 °C, respectively.

The crystal structures were examined by x-ray diffractometer using Cu

$K_\alpha$ radiation at room temperature. The magnetic measurements were carried out with a Quantum Design superconducting quantum interference device (SQUID) MPMS system (2 ≤T≤400 K, 0 ≤ H ≤5 T). Both zero-field-cooling (ZFC) and field-cooling (FC) data were recorded. The resistance was measured by the standard four-probe method from 25 to 300 K.

Fig.1 shows the x-ray diffraction (XRD) pattern of $La_{0.9}Te_{0.1}MnO_3$ samples of (A) 800 °C, (B) 900 °C, (C) 1000 °C-sintered by a citrate-gel technique and (D) obtained through the conventional solid-state reaction method. The powder x-ray diffraction at room temperature shows that all samples are single phase with no detectable secondary phases and the samples have a rhombohedral structure with the space group $R\bar{3}C$. The average grain size $D_{hkl}$ is estimated through the classical Scherrer formula [20] $D_{hkl} = k\lambda / B\cos\theta$, where $D_{hkl}$ is the grain size derived from the (0 2 4) peak of the XRD profiles, $k$ is a constant (shape factor ~ 0.9), $\theta$ is the angle of the diffraction, $B$ is the difference of the full width at half-maximum (FWHM) of the peak between the sample and the standard $SiO_2$ used to calibrate the intrinsic width associated to the instruments, and $\lambda$ is the wavelength of X-ray. The grain size obtained from the Scherrer formula is thus estimated as 40 nm, 53 nm, 82 nm, and 120 nm for samples A, B, C, and D, respectively. As it can be clearly seen from the inset of Fig.1, the grain size increases with increasing $T_s$ and the diffraction peak (0 2 4) shifts towards the lower angle from sample A to D, which is consistent with many results reported elsewhere [11,16]. The structural parameters of the samples are refined by the standard Rietveld technique [21] and the fitting between the experimental spectra and the calculated values is quite well based on the consideration of lower $R_P$ values as shown in Table 1. It shows that the lattice parameters of $La_{0.9}Te_{0.1}MnO_3$ samples vary monotonously with increasing the sintered temperature $T_s$. The Mn-O-Mn bond angle decreases and the Mn-O bond length increases with the increase of the particle size.

Fig.2 shows the temperature dependence of magnetization M of $La_{0.9}Te_{0.1}MnO_3$ under both zero-field-cooled (ZFC) and field-cooled (FC) modes at H = 0.1 T for all samples. The Curie temperature $T_C$ (defined as the one corresponding to the peak of $dM/dT$ in the M vs. T curve) is 268, 253, 240 and 239K for samples A, B, C and D, respectively. Obviously, the Curie temperature decreases with the increase of $T_s$ from sample A to C. $T_C$ of sample D is almost same as that of sample C. It demonstrates that there may exist a critical value of the grain size. Below this critical value, the Curie temperature of the sample may be affected strongly by the grain size. Whereas beyond this critical value, the grain size does not almost affects $T_C$. Here, for the studied sample $La_{0.9}Te_{0.1}MnO_3$, the critical value of the grain size may be about 80 nm. We suggest that the $T_C$ reduction should be attributed to the weakening of double exchange (DE) interaction because of the decrease of the bandwidth and the mobility of $e_g$ electrons due to the increase of Mn-O bond length and the decrease of Mn-O-Mn bond angle. Additionally, it is surprising that the magnetization M decreases at low temperatures with the increase of grain size, which is exact contrary to those obtained by Sánchez et al. [11], in which both magnetization and $T_C$ increase with increasing grain size of $La_{0.67}Ca_{0.33}MnO_3$.

The magnetization as a function of the applied magnetic field at 5 K is shown in Fig.3. It shows that, for all samples, the magnetization reaches saturation at about 1T and keeps constant up to 5T, which was considered as a result of the rotation of the magnetic domain under the action of applied magnetic field. Additionally, similar to the result shown in Fig.2, the magnetization M decreases at low temperatures with increase grain size. According to DE theory, the magnetization M is given by [22] $M \propto xb/4|J|S^2$, where b is the transfer integral between neighboring Mn ions, S the ionic spin, J the intra-atomic exchange integral, and x the doping level of the compound. In the present case, two possible factors influencing M are x and b. The oxygen content of the samples was determined by a redox (oxidation reduction) titration. The detailed method to determine the oxygen content of samples will be reported in elsewhere [23]. Only slightly surplus oxygen has been observed for our samples and the oxygen content almost stabilize around 3.01. So the factor of the doping level can be excluded. In addition, the change of lattice parameters, especially the increase of Mn-O bond length ($d_{Mn-O}$) with grain size, needs to be taken into account. The increase of $d_{Mn-O}$ is to decrease the overlap between the neighboring orbits of Mn ions and the adjacent O ions, thereby reducing the exchange integral b, which described electron hopping between Mn sites. Evidently, the magnetization decreases with the increase of grain size can be attributed to the increase of the Mn-O bond length $d_{Mn-O}$.

Fig.4 shows the $\rho - T$ curves at zero field and under an applied field H = 0.5T, together with the corresponding MR-T curves for the samples sintered at different temperatures, where magnetoresistance (MR) is defined as $\Delta\rho/\rho_0 = [(\rho_0 - \rho_H)/\rho_0] \times 100\%$. For sample A, it shows that there exists an insulator-metal (I-M) transition at $T_{P2}$ (= 222K) which is well below its Curie temperature $T_C$ (= 268K). For sample B, it shows that there exists an I-M transition at $T_{P1}$ (= 252 K) which is close to its Curie temperature $T_C$ (= 253K). In addition, there exists a noticeable bump at $T_{P2}$ (= 224 K) below $T_{P1}$. The value of $T_{P1}$ and $T_{P2}$ are listed in Table 2. For samples C and D, I-M transition occurs at almost the same temperature $T_{P1}$ (245K and 246K, respectively). In addition, there exists a shoulder at $T_{P2}$ (221K, and 223K for sample C, and D, respectively). Compared with sample A, I-M transition at $T_{P1}$ becomes more obvious and I-M transition at $T_{P2}$ becomes weaker. It shows that the height of the resistivity peak at $T_{P2}$ decrease with increasing grain size, which is tightly related to grain boundaries (GBs). Here, GBs can act as tunneling barriers for spin-dependent transport. This variation behavior of double ρ peaks is presumably related to the increase of both the height and width of tunneling barriers with decreasing the grain size [24]. Therefore, the height of the resistivity peak at $T_{P2}$ increases from sample D to B and finally exhibits only one resistivity peak in the $\rho-T$ curve for sample A. Different from the origin of the I-M transition at $T_{P1}$, the resistivity peak at $T_{P2}$ is believed to reflect the spin-dependent interfacial tunneling due to the difference in magnetic order between surface and core [14]. Moreover, Fig.4 manifests that resistivity of sample A increases about one order of magnitude compared to that of samples B, C, and D. Although the resistivity order of magnitude is the same for samples B, C, and D, $\rho$ decreases slightly with increasing grain size. We consider that the increase of $\rho$ with the reduction of grain size is mainly related to the increase of both the height and width of tunneling barriers with decreasing the grain size. The experimental data measured at applied field of 0.5 T for all samples in the temperature range of 30-300K are also recorded. For samples B, C, and D, it can be seen from the $\rho-T$ curves in Fig.4 that the applied field suppressed the resistivity peak at $T_{P1}$ significantly and the resistivity peak shifts towards higher temperatures. This suggests that the external magnetic field facilitates the hopping of $e_g$ between the neighboring Mn ions, which agrees with the DE model. However, for the second I-M transition at $T_{P2}$, it is worth noting that for all samples, the position of the resistivity peak at $T_{P2}$ almost does not change under the applied magnetic field. The difference in the response of the resistivity peak at $T_{P1}$ and $T_{P2}$ for the applied field in samples B, C, and D indicates again that they may have different origins. It can also be seen from the temperature dependence of MR for different samples shown in Fig.4. Evidently, for sample A, it shows that MR effect below $T_C$ increases with decreasing temperature, which is a key feature of the extrinsic IMR. For samples B, C, and D, both the intrinsic CMR effect showing corresponding peak in the vicinity of $T_{P1}$ and the extrinsic IMR below $T_C$ can be simultaneously observed, which is just corresponding to the intrinsic transport properties and spin-dependent tunneling transport, as can be seen clearly from the double-peak-type $\rho - T$ curves in Figs.4 (b), 4(c) and 4(d). Additionally, for sample A, the extrinsic IMR becomes so important that the intrinsic counterpart influence is negligible and the MR ratio around 30K is obtained as high as 33%. For samples B, C, and D, as it can be seen clearly from Fig.4, the intrinsic CMR effect becomes more prominent with increasing grain size, e.g., the MR value of 7%, 18%, and 23% around $T_{P1}$ for samples B, C, and D, respectively. Whereas the extrinsic IMR effect at low temperatures becomes smaller with increasing grain size, e.g., the MR value of 26%, 20%, and 14% around 30K for samples B, C, and D, respectively. Evidently, the MR effect can be effectively tuned in $La_{0.9}Te_{0.1}MnO_3$ by changing grain size.

In summary, we have studied the structural, magnetic and transport properties of $La_{0.9}Te_{0.1}MnO_3$ with different grain sizes. The results show that the Mn-O-Mn bond angle decreases and the Mn-O bond length increases with the increase of grain size. The magnetic measurement exhibits that both magnetization and the Curie temperature $T_C$ decrease with increasing grain size is observed, which is mainly proposed to be related to the increase of the Mn-O bond length $d_{Mn-O}$. For sample A with grain size of 40 nm, only the extrinsic IMR below $T_C$ is observed. The results reveal that both the intrinsic CMR and extrinsic IMR can be effectively tuned in $La_{0.9}Te_{0.1}MnO_3$ by changing grain size.

**ACKNOWLEDGMENTS**

This work was supported by the National Key Research under contract No.001CB610604, and the National Nature Science Foundation of China under contract No.10174085, Anhui Province NSF Grant No.03046201 and the Fundamental Bureau of Chinese Academy of Sciences.

**68** (1996) 2291.

**Table 1. Refined structural parameters of $La_{0.9}Te_{0.1}MnO_3$ at room temperature. The space group is $R\bar{3}C$.**

| Sample | a (Å) | c (Å) | V (Å$^3$) | d$_{Mn-O}$ (Å) | θ$_{Mn-O-Mn}$ (º) | R$_p$ (%) |
|---|---|---|---|---|---|---|
| A | 5.5163 | 13.3482 | 351.9913 | 1.9586 | 164.71 | 9.81 |
| B | 5.5214 | 13.3531 | 352.4064 | 1.9602 | 164.37 | 7.66 |
| C | 5.5223 | 13.3561 | 353.0042 | 1.9637 | 163.85 | 8.72 |
| D | 5.5241 | 13.3572 | 353.0103 | 1.9644 | 163.83 | 8.03 |

**Figure captions**

Fig.1. X-ray diffraction (XRD) pattern of $La_{0.9}Te_{0.1}MnO_3$ samples of (A) 800 °C, (B) 900 °C, (C) 1000 °C-sintered by a citrate-gel technique and (D) the bulk polycrystalline obtained through the conventional solid-state reaction method. The inset shows the width of the diffraction peak (0 2 4) for different sintering temperatures $T_s$.

Fig.2 Magnetization as a function of temperature M (T) for samples A, B, C, and D under the field-cooled (FC) and zero-field-cooled (ZFC) modes denoted as the filled and open symbols, respectively.

Fig.3. Field dependence of the magnetization M(H) for samples A, B, C and D at 5 K.

Fig.4. (a), (b), and (c) Temperature dependence of resistivity ρ(T) at zero field (solid lines) and under an applied field of 0.5T (dashed lines), and the corresponding temperature dependence of MR for samples A, B, and C, respectively; (d) the experimental data of ρ vs T and MR vs T for sample D.

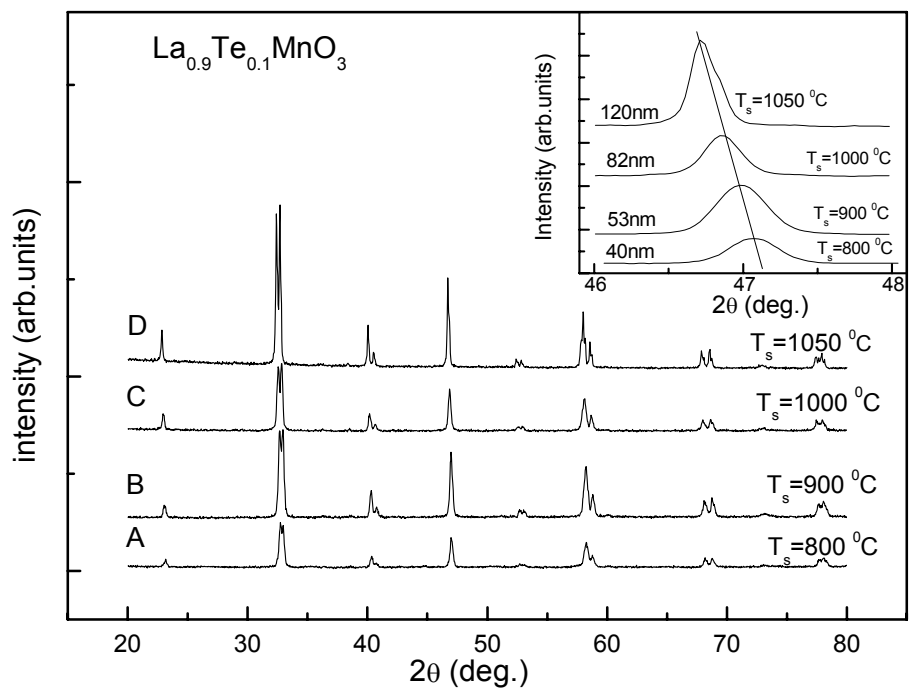

Fig.1 J. Yang et al.

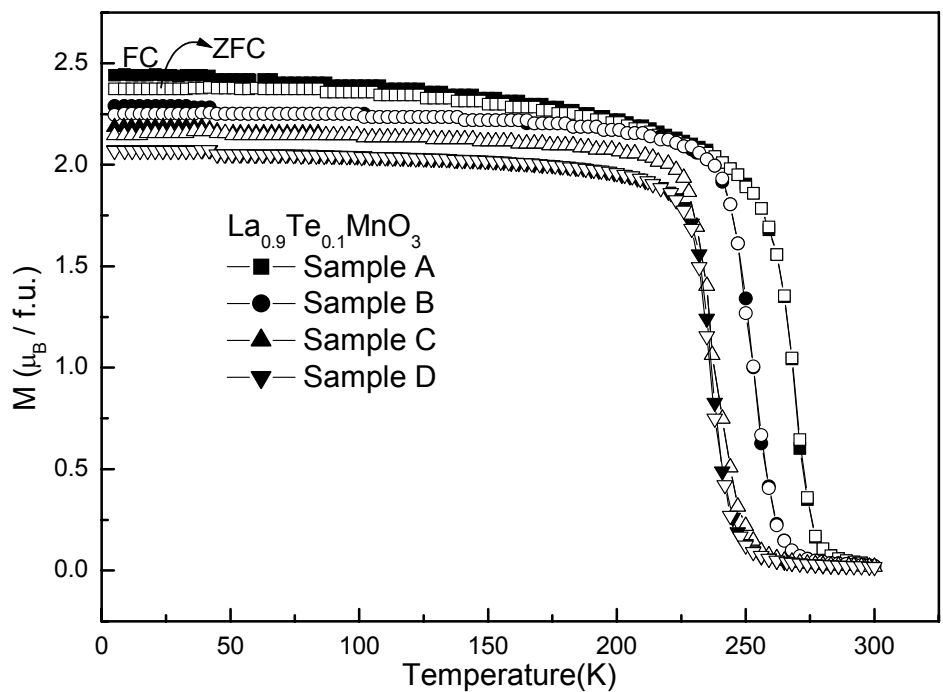

Fig.2 J. Yang et al.

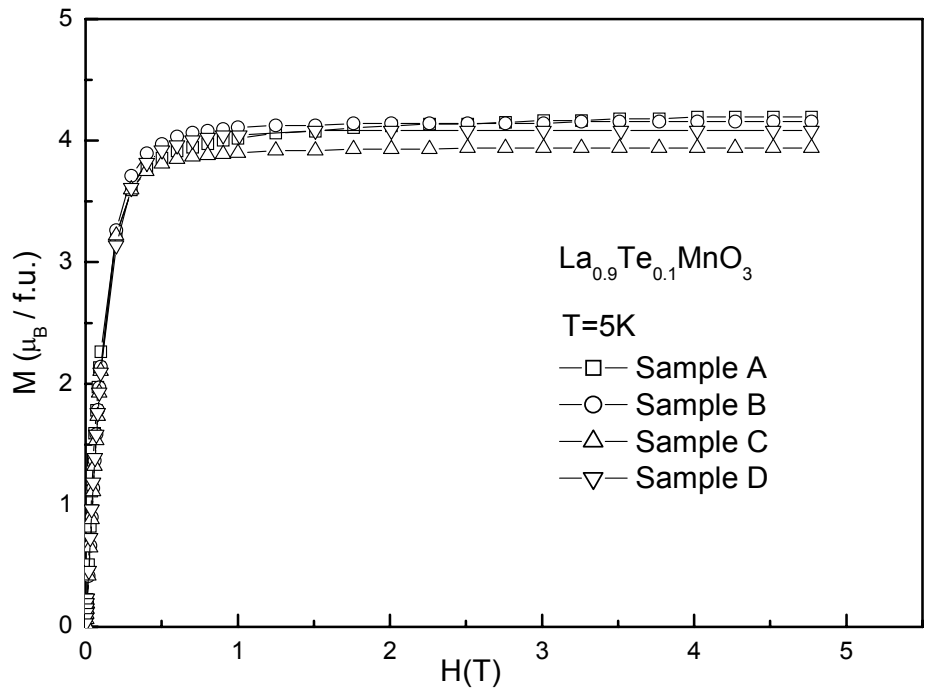

Fig.3 J. Yang et al.

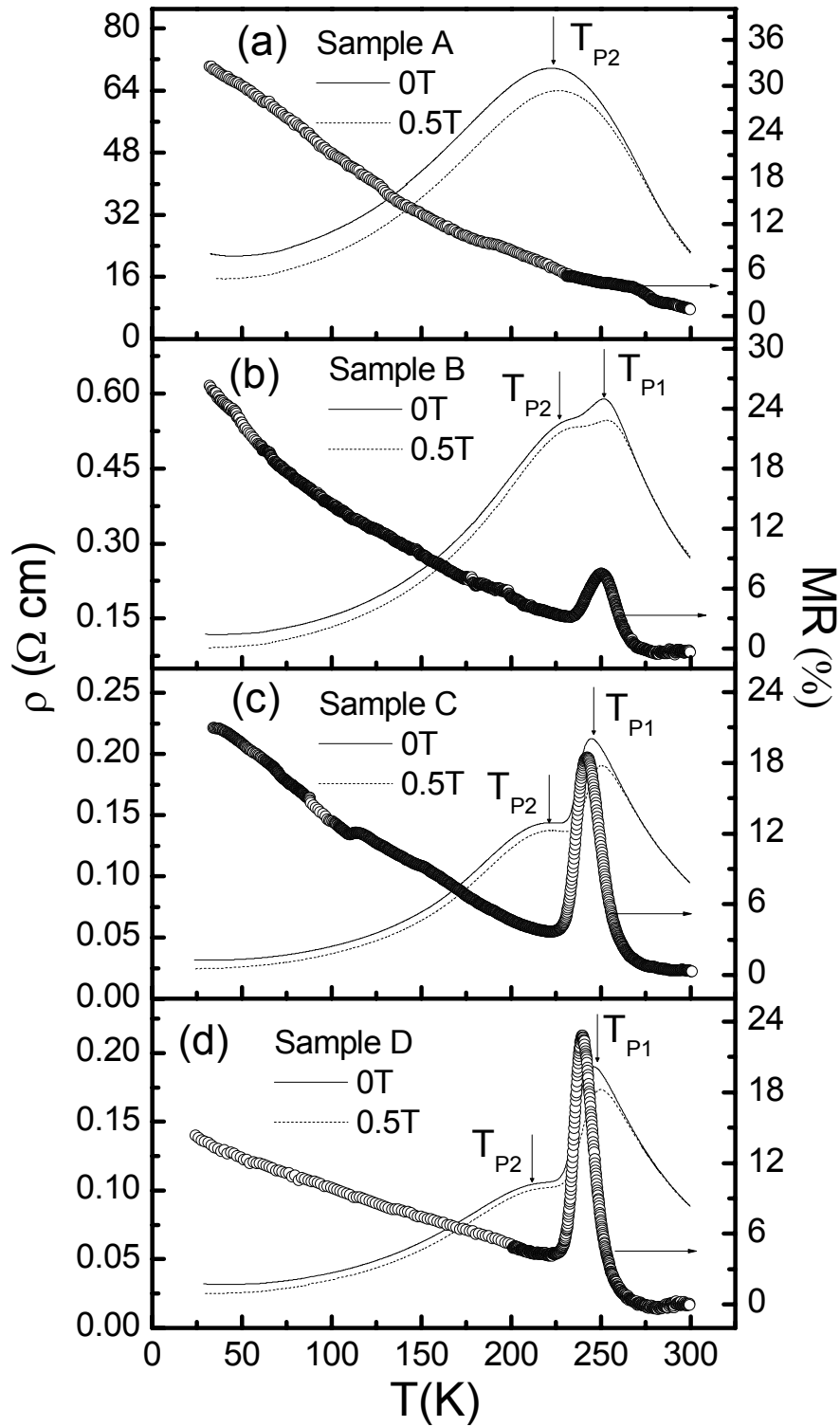

Fig.4 J. Yang et al.